\documentclass[aps, 10pt, prl,
notitlepage, twocolumn, superscriptaddress,
nofootinbib]{revtex4-1}

\usepackage{amsmath}
\usepackage{amssymb}
\usepackage{amsfonts}
\usepackage[utf8]{inputenc}
\usepackage[T1]{fontenc}
\usepackage{mathrsfs}
\usepackage{anyfontsize}

\usepackage[linktocpage,breaklinks]{hyperref}
\usepackage[usenames,dvipsnames]{xcolor}

\usepackage{txfonts}
\usepackage{tensor}
\usepackage{bm}

\usepackage{graphicx}
\usepackage{epsfig}
\usepackage{epstopdf}

\usepackage{natbib}
\usepackage{hyperref}

\hypersetup{colorlinks=true, citecolor=Plum, linkcolor=Plum,
urlcolor=Plum}

\newcommand{\dd}{{\rm d}}

\newcommand{\vp}{\varphi}
\newcommand{\del}{\partial}
\newcommand{\calg}{\mathscr G}

\newcommand{\calv}{\mathscr V}
\newcommand{\msun}{$_{\odot}$}

\begin{document}
\title{Spontaneous scalarization of black holes and compact stars \\
from a Gauss--Bonnet coupling}

\begin{abstract}
We identify a class of scalar-tensor theories with coupling between
the scalar and the Gauss--Bonnet invariant that exhibit spontaneous
scalarization for both black holes and compact stars.
In particular, these theories formally admit all of the stationary
solutions of general relativity, but these are not dynamically preferred
if certain conditions are satisfied. Remarkably, black holes exhibit
scalarization if their mass lies within one of many narrow bands.
We find evidence that scalarization can occur in neutron stars as well.
\end{abstract}

\author{Hector O. Silva}
\email{hector.okadadasilva@montana.edu}
\affiliation{Department of Physics and Astronomy,
The University of Mississippi, University, MS 38677, USA}
\affiliation{eXtreme Gravity Institute,
Department of Physics, Montana State University, Bozeman, MT 59717 USA}

\author{Jeremy Sakstein}
\email{sakstein@physics.upenn.edu}
\affiliation{Center for Particle Cosmology,
Department of Physics and Astronomy,
University of Pennsylvania,
209 S. 33rd St., Philadelphia, PA 19104, USA}

\author{Leonardo Gualtieri}
\email{leonardo.gualtieri@roma1.infn.it}
\affiliation{Dipartimento di Fisica ``Sapienza''
Universit\`a di Roma \& Sezione INFN Roma1,
Piazzale Aldo Moro 5, 00185, Roma, Italy}

\author{Thomas P. Sotiriou}
\email{thomas.sotiriou@nottingham.ac.uk}
\affiliation{School of Mathematical Sciences,
University of Nottingham, University Park, Nottingham, NG7 2RD, UK}
\affiliation{School of Physics and Astronomy,
University of Nottingham, University Park, Nottingham, NG7 2RD, UK}

\author{Emanuele Berti}
\email{eberti@olemiss.edu}
\affiliation{Department of Physics and Astronomy,
The University of Mississippi, University, MS 38677, USA}

\date{{\today}}

\maketitle

\textit{Introduction.}
Gravitational wave observations~\cite{Abbott:2016blz,Abbott:2016nmj,TheLIGOScientific:2016pea,Abbott:2017vtc,Abbott:2017oio,Abbott:2017gyy,TheLIGOScientific:2017qsa}
allow us to probe the structure of
black holes (BHs) with unprecedented accuracy. Hence, they can reveal the
existence of new fundamental scalar
fields~\cite{Berti:2015itd,Sotiriou:2015lxa}, provided that they leave an
imprint on BHs.
However, no-hair theorems
(see~\cite{Sotiriou:2015pka,Herdeiro:2015waa} for reviews) dictate
that conventional scalar-tensor theories will have the same
stationary, asymptotically flat BH solutions as general relativity
(GR) \cite{Hawking:1972qk,Bekenstein:1971hc,Sotiriou:2011dz}.
In spherical symmetry \cite{Hui:2012qt} and slow
rotation~\cite{Sotiriou:2013qea,Sotiriou:2014pfa}, this result extends
to generalized scalar-tensor theories, i.e.~theories that exhibit
derivative self-interactions and derivative couplings between the scalar
and curvature invariants, provided that the scalar respects shift symmetry.

One could still detect scalars in these theories through the imprint
they leave when they are
excited~\cite{Barausse:2008xv,Arvanitaki:2009fg}. One can also
circumvent no-hair theorems by violating some of their
assumptions~\cite{Cardoso:2013fwa,Cardoso:2013opa,Babichev:2013cya,Herdeiro:2014goa}.
No-hair theorems also help single out particularly interesting
theories that have hairy BHs. A well-studied example is the action
\begin{align}
S = \frac{1}{2}\int\dd^4 x\sqrt{-g}
\left[R-\frac{1}{2}\nabla_\alpha\vp\nabla^\alpha\vp
+ f(\vp)\calg\right]
+ S_{\rm m}[g_{\mu\nu},\psi]\,, 
\label{eq:action}
\end{align}
where $\calg \equiv R_{\mu\nu\rho\sigma} R^{\mu\nu\rho\sigma} -
4R_{\mu\nu} R^{\mu\nu} + R^2$ is the Gauss-Bonnet invariant.
We use geometrical units with $c=8\pi G=1$ and the mostly plus metric
signature. The scalar field $\vp$ is coupled to $\calg$, which has
dimensions of length$^{-4}$ ($\equiv L^{-4}$), through a function
$f(\vp)$, with dimensions $L^2$.
The matter fields $\psi$ are minimally coupled to the metric
$g_{\mu\nu}$ through the action $S_{\rm m}$. We will refer to this
class of theories as scalar-Gauss-Bonnet (sGB) gravity. When $f$ is
exponential the theory is well-known to admit hairy
BHs~\cite{Kanti:1995vq}, whereas a linear $f$ yields the only
shift-symmetric theory with second-order field equations that exhibits
BH hair~\cite{Sotiriou:2013qea,Sotiriou:2014pfa} (despite the no-hair
theorem of~\cite{Hui:2012qt}).

The main purpose of this Letter is to demonstrate that a new subclass
of theories, contained in~\eqref{eq:action}, exhibits a particularly
interesting phenomenon: BH spontaneous scalarization.  As we
demonstrate below, this subclass of theories generically admits
solutions where the scalar field is constant and the metric satisfies
Einstein's equations. However, under certain conditions, these
solutions are unstable, and solutions where the scalar field is
nontrivial are dynamically preferred. This leads to hairy BHs only
when the BH mass lies within certain ranges. Compact stars in these
theories also exhibit spontaneous scalarization. The mechanism
resembles that proposed by Damour and Esposito-Far\`ese
\cite{Damour:1993hw}, where there is a coupling between $\varphi$ and
the trace of the stress-energy tensor, $T$.
However, there are important differences -- most notably the fact that the
effect is present for BHs as well.

\textit{A no-hair theorem in sGB and how to evade it.}
We start by identifying the class of theories in question.
Varying~\eqref{eq:action} with respect to $\vp$ and $g_{\mu\nu}$ yields
\begin{subequations}
\begin{align}
&\Box \vp  = - f_{,\vp} \calg\,,
\label{eq:eom_scalar}
\\
&R_{\mu\nu} - \frac{1}{2} g_{\mu\nu} R = T_{\mu\nu}\,.
\label{eq:eom_metric}
\end{align}
\end{subequations}
Here $T_{\mu\nu}$ is the sum of the matter stress-energy tensor
$T^{\rm m}_{\mu\nu} \equiv -(2 / \sqrt{-g}) (\delta S_{\rm m}/\delta
g^{\mu\nu})$, plus a contribution coming from the variation of the
$\varphi$-dependent part of the action with respect to the metric (see
e.g.~\cite{Kanti:1995vq}).

Eq.~\eqref{eq:eom_scalar} does not admit $\vp =$ constant solutions,
unless
\begin{equation}
f_{,\vp}(\vp_0) = 0\,,
\label{eq:existence}
\end{equation}
for some constant $\vp_0$. We consider Eq.~\eqref{eq:existence} as an
{\it existence condition} for GR solutions and focus on theories
that satisfy it. This excludes the widely studied class of dilatonic
theories where $f \sim \exp(\vp)$ and the shift-symmetric $f \sim \vp$
theory discussed above~\cite{Kanti:1995vq,Sotiriou:2013qea,Sotiriou:2014pfa}.

Focus now on BH solutions that are asymptotically flat and stationary.
These admit a Killing vector $\xi^\mu$ that is timelike at infinity
and acts as a generator of the event horizon.  Assuming that $\varphi$
respects stationarity, $\xi^\mu\nabla_\mu \vp=0$.
Multiplying Eq.~\eqref{eq:eom_scalar} by $f_{,\vp}$ and integrating
over a volume $\calv$ yields
\begin{equation}
\int_\calv \dd^4x\sqrt{-g} \left[f_{,\vp}\Box\vp
+f^2_{,\vp}(\vp) \calg\right]=0\,.
\end{equation}
Integrating by parts and using the divergence theorem, we obtain
\begin{align}
&\int_\calv \dd^4x\sqrt{-g} \left[f_{,\vp\vp}\nabla^\mu \vp \nabla_\mu\vp -f^2_{,\vp}(\vp) \calg\right] \nonumber \\
&=
\int_{\partial\calv}\dd^3x\sqrt{|h|} f_{,\vp} n^\mu \nabla_\mu\vp\,,
\label{proof}
\end{align}
where $\partial \calv$ is the boundary of $\calv$ and
$n^\mu$ is the normal to the boundary.
We choose $\calv$ such that it is bounded by the BH
horizon, two partial Cauchy surfaces, and spatial infinity.
The contribution of the boundary term on the
right-hand side vanishes. The horizon contribution vanishes by
symmetry, as the normal to the horizon is $\xi^\mu$ and
the stationarity condition holds;
the contribution of the boundary at infinity vanishes because of
asymptotic flatness. The contributions of the Cauchy surfaces exactly
cancel each other, as they can be generated by an isometry. Hence the
integral in the first line of Eq.~\eqref{proof} must vanish as
well. With our signature, $\nabla^\mu \vp \nabla_\mu\vp$ is positive
in the BH exterior.  Indeed, whenever
\begin{equation}
f_{,\vp\vp}\, \calg<0
\label{eq:stability}
\end{equation}
the whole integrand is sign definite and must vanish at every point in
$\calv$. The same conditions imply that the two terms of the integrand
have the same sign and hence must vanish separately. This can only be
achieved if $\vp=\vp_0$.

The above can be considered as a no-hair theorem for stationary,
asymptotically flat BHs in theories that satisfy the conditions of
Eqs.~\eqref{eq:existence} and \eqref{eq:stability}. The former is clearly
an existence condition for GR solutions. To understand the latter, it is
helpful to linearize Eq.~\eqref{eq:eom_scalar} around
$\vp = \vp_0$,
\begin{equation}
\left[\Box + f_{,\vp\vp}(\vp_0)\, \calg\right]\delta\vp = 0\,.
\label{eq:perturbeq}
\end{equation}
The term $\,-f_{,\vp\vp}\, \calg$ acts as an effective mass
$m^2_{\rm eff}$ for the perturbations $\delta\vp$.
Theories for which this effective mass is negative can
evade the theorem above.
There is a direct analogy between the proof presented here and the
no-hair theorem proof of~\cite{Sotiriou:2011dz} for scalar-tensor
theories with self-interactions.

This no-hair theorem identifies theories that can lead to interesting
phenomenology in the strong-field regime: they must satisfy
condition~\eqref{eq:existence} but violate
condition~\eqref{eq:stability}. A negative effective mass is expected
to trigger a tachyonic instability, which can lead to the development
of {\it scalar hair}. This is analogous to spontaneous scalarization
for neutron stars (NSs) in standard scalar-tensor
theories~\cite{Damour:1993hw}. Scalarization was also shown to be
possible for BHs if they are surrounded by
matter~\cite{Cardoso:2013fwa,Cardoso:2013opa}.

\textit{Quadratic scalar-Gauss-Bonnet gravity.}
The simplest coupling function which satisfies
Eq.~\eqref{eq:existence} and can violate Eq.~\eqref{eq:stability} is
\begin{equation}
f= \eta \vp^2 / 8\,,
\label{quadraticf}
\end{equation}
where $\eta$ is a parameter with dimensions $L^2$. Hereafter, we will
focus on this theory, and we will call it quadratic sGB (qsGB)
gravity. If $f$ satisfies the condition \eqref{eq:existence} and is
well behaved around $\varphi_0$, then it admits the expansion
$f(\varphi)=f(\varphi_0)+f_{,\varphi\varphi}(\varphi_0)(\varphi-\varphi_0)^2/2+\ldots$
The first term in this expansion does not contribute to the field equations
because $\calg$ is a total divergence. Moreover, the kinetic term of the
action is shift-symmetric. So, the field redefinition
$\varphi\to \varphi-\varphi_0$ can reduce the quadratic expansion of any
theory to qsGB.

qsGB gravity has several other interesting features. It leads to a field
equation for $\varphi$ that is linear in $\varphi$. This will be
particularly convenient when studying the zero-backreaction limit below.
Additionally, the theory exhibits $\varphi\to-\varphi$ symmetry. This
is important in a field theory context. It prevents the term
$\varphi \calg$, which inevitably leads to BH hair
\cite{Sotiriou:2013qea,Sotiriou:2014pfa}, from appearing in the
action. Note also that $\varphi$ does not need to play any role in
late-time cosmology, hence current weak-field and gravitational
wave constraints are very
weak~\cite{EspositoFarese:2004cc,Sotiriou:2006pq,Sakstein:2017bws,Sakstein:2017xjx}.

We focus on spherically symmetric solutions that describe either BHs
or compact stars and demonstrate that spontaneous scalarization can
take place.  We first consider the scalar on a GR background and show
that there is an instability associated with spontaneous
scalarization. We then verify our results by looking at
non-perturbative solutions. We call the solution with a non-trivial
scalar configuration the {\it scalarized solution}. We focus on
solutions that share the same asymptotics with the GR solution,
including the asymptotic value of $\varphi$, $\vp_\infty$. For
simplicity, we impose $\vp_\infty=0$, but this choice does not
crucially affect our results.

\textit{Tachyonic instability: a zero-backreaction analysis.}
We first consider the limit where backreaction from the metric can be
neglected; i.e.,~we focus on the scalar field equation,
Eq.~\eqref{eq:perturbeq}, on a fixed background.
The effective mass of the perturbation $\delta\vp$ is
$m^2_{\rm eff} = -f_{\vp\vp}\calg = - \eta\,\calg / 4$, therefore
tachyonic instability should be possible for $\eta>0$. On a static,
spherically symmetric background spacetime
$\dd s^2 = -a(r)\dd t^2 + b(r) \dd r^2 + r^2 \dd\Omega$,
Eq.~\eqref{eq:perturbeq} can be written as
\begin{equation}
-\frac{\del^2 \sigma}{\del t^2}
+\frac{\del^2 \sigma}{\del r_{\ast}^2}
= V_{\rm eff}\, \sigma\,,
\label{eq:waveeq_decoupl}
\end{equation}
where $\delta \vp = \sigma(t,r) Y_{\ell m}(\theta,\phi)/r$, $Y_{\ell m}$ are standard spherical harmonics, $\dd r/\dd
r_{\ast} \equiv \sqrt{a/b}$ and the effective potential $V_{\rm eff}$ is:
\begin{equation}
V_{\rm eff}
\equiv
a\left[
\frac{\ell(\ell + 1)}{r^2} + \frac{1}{2ra}\frac{\dd (a b^{-1})}{\dd r}
- \frac{\eta\,\calg}{4}
\right]\,.
\label{eq:effective_potential}
\end{equation}

\begin{figure}[t]
\includegraphics[width=0.5\textwidth]{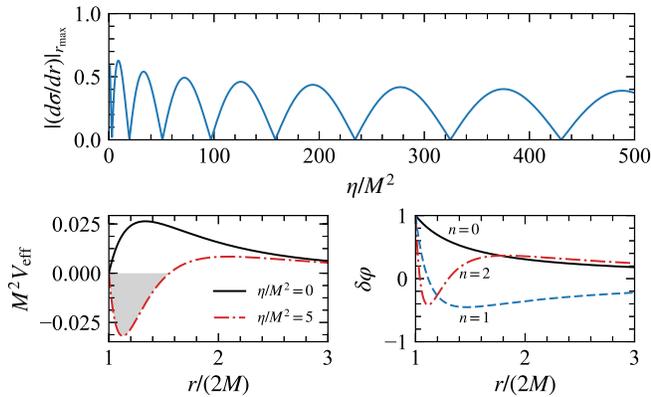}
\caption{{\it Scalar field in the decoupling limit}. Results of the
  numerical integration of the decoupled scalar field
  equation~\eqref{eq:waveeq_decoupl}, assuming $\ell=0$ and a
  Schwarzschild background.  Top panel: asymptotic value of the scalar
  field as a function of $\eta / M^2$. Cusps correspond to scalarized
  solutions.  Bottom-left panel: effective potential $V_{\rm eff}$ for
  $\eta / M^2 = 0$ and $5$. In the latter case $V_{\rm eff}$ develops
  a negative region and it can support bound states.  Bottom-right
  panel: radial profiles of $\delta \vp$ for the first three
  scalarized solutions, corresponding to $\eta /M^2 = 2.902$, $19.50$
  and $50.93$. These profiles have $0$, $1$ and $2$ nodes,
  respectively.}
\label{fig:decoupl}
\end{figure}

In order to find whether scalarized solutions of the decoupled field
equation~\eqref{eq:waveeq_decoupl} exist, we have performed a
numerical integration, assuming a Schwarzschild background and
monopolar perturbations. We have found that the equation admits a
non-trivial solution with $\varphi_\infty=0$ for a discrete spectrum
of values of the coupling parameter ($\eta /M^2 = 2.902$, $19.50$,
$50.93$, $\dots$). These results are summarized in
Fig.~\ref{fig:decoupl}, where we show the quantity $d\sigma/dr$
computed at some extraction radius $r_{\rm max}\gg M$ (namely
$r_{\rm max}=200\,M$), as a function of $\eta/M^2$. For $r\gg M$,
$\delta\vp\sim\delta\vp_\infty+O(r^{-1})$, thus
$\delta\vp_\infty \sim d\sigma/dr(r\rightarrow\infty)$. The scalarized
solutions correspond to the cusps in the top panel of
Fig.~\ref{fig:decoupl}. These solutions can be characterized by an
order number $n=0,1,\dots$, which is also the number of nodes of the
radial profile of $\delta\vp(r)$ (bottom-right panel of
Fig.~\ref{fig:decoupl}).

\textit{Scalarized black holes in qsGB gravity.}
We now consider BH solutions obtained by integrating the full set of
equations~\eqref{eq:eom_scalar} and \eqref{eq:eom_metric}.  We search
for static, spherically symmetric solutions, i.e.
$a=a(r),\,b=b(r),\,\vp=\vp(r)$. We define $\Gamma=\log a$,
$\Lambda=\log b$, as in \cite{Kanti:1995vq}. The field equations can
be cast as three coupled ordinary differential equations for $\Gamma$,
$\Lambda$ and $\vp$. Since these equations are not particularly
illuminating, we do not present them here.

The equation for $\Lambda$ can be integrated
algebraically~\cite{Kanti:1995vq,Sotiriou:2013qea,Sotiriou:2014pfa}:
\begin{equation}
e^{\Lambda} = \frac{-A + \delta \sqrt{A^2 - 4 B}}{2}, \quad \delta = \pm 1\,,
\label{eq:alg_lambda}
\end{equation}
where $A = (1/4)r^2 \vp'^2 - (r + \eta \vp \vp'/2)\Gamma' - 1$ and
$B = (3/2)\Gamma' \vp' \vp$.
In BH solutions $\exp(-\Lambda),\, \exp(\Gamma) \to \infty$ at the event
horizon $r_{\rm h}$, and this implies $\delta=1$~\cite{Kanti:1995vq}.
Replacing Eq.~\eqref{eq:alg_lambda} in the remaining equations, we are left
with two differential equations for $\Gamma$ and $\vp$. A near-horizon
expansion of the field equations shows that $\vp''_{\rm h}=\vp''(r=r_{\rm h})$
is finite if
\begin{equation}
\vp'_{\rm h} = \frac{r_{\rm h}}{\eta\vp_{\rm h}}
\left(
- 1 + \xi \sqrt{1 -{6 \eta^2 \vp_{\rm h}^2}/{r_{\rm h}^4}}
\right)\,,
\label{eq:finite_phi}
\\
\end{equation}
where $\xi = \pm 1$. The $\xi = -1$ branch does not result in a BH
solution, as discussed in~\cite{Kanti:1995vq} for the exponential
coupling. Therefore, regularity on the horizon requires
\begin{equation}
r_{\rm h}^4 - 6 \eta^2 \vp_{\rm h}^2 \geq 0\,.
\label{eq:ineq}
\end{equation}
Eq.~\eqref{eq:ineq}
defines a region in the ($r_{\rm h}, \vp_{\rm h}$) plane within
which BH solutions with a regular (real) scalar field configuration
exist.

The value of the scalar field at the horizon is bound in the range
$0\le \vp_{\rm h}\le\vp_{\rm h}^{\rm max}=r_{\rm h}^2/(\sqrt{6}\eta)$.
We do not consider solutions with $\vp_{\rm h}<0$ because qsGB gravity
is invariant under $\vp\to-\vp$. The field equations are invariant
under the rescalings $r_{\rm h}\to r_{\rm h}/l$, $M\to M/l$,
$\eta\to\eta/l^2$, corresponding to a freedom in choosing length
units. BH solutions are then characterized by dimensionless quantities
such as $\eta/M^2$ and $\eta/r_{\rm h}^2$.

\begin{figure}[t]
\includegraphics[width=0.5\textwidth]{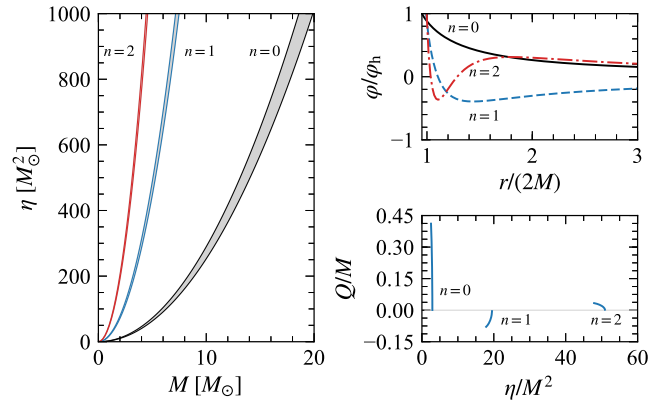}
\caption{{\it Spontaneous scalarization of black holes.} Left: the regions
in the $\eta-M$ (in solar mass units) space
where scalarized BHs exist. The solutions belonging to each band are
characterized by the number of nodes of the
scalar field radial profile.  We only show the first three
scalarization regions, but our numerical analysis suggests an
infinite number of them.
Top-right: the scalar field profiles for sample
BH solutions in each of the first three bands.
Bottom-right: normalized scalar charge $Q/M$ as a function of
$\eta / M^2$. The most charged BHs belong to the $n = 0$ band.}
\label{fig:bh}
\end{figure}

For each value of $\eta/M^2$ we have numerically solved the field
equations, with $\vp_{\rm h}$ in the range $[0,\vp_{\rm}^{\rm max}]$
and the other boundary conditions fixed from the requirement of
regularity at the horizon. We have then extracted the scalar
quantities characterizing the solution -- the mass $M$, the scalar
charge $Q$, and the asymptotic value of the scalar field $\vp_\infty$
-- from the asymptotic
expansions~\cite{Mignemi:1992nt,Kanti:1995vq,Sotiriou:2014pfa}:
\begin{align}
e^{\Gamma} &= 1 - 2 M/r + Q^2 M/(12 r^2)\,, \\
\vp &= \vp_0 + Q/r + Q M/r^2 + (32 Q M^2 - Q^3)/(24 r^3)\,.
\label{asyntexp}
\end{align}
While the Schwarzschild solution ($\vp_{\rm h}=0$, $\vp_0=0$) is
allowed for any value of $\eta$, a solution with $\vp_{\rm h}\neq0$,
$\vp_\infty=0$ only exists when $\eta/M^2$ belongs to a set of
``scalarization bands'', i.e. $[2.53,2.89]$, $[17.86,19.50]$,
$[47.90,50.92]$, etc. The right-end values of these bands correspond
to the eigenvalues of $\eta/M^2$ found by solving the linear equation
of the scalar field on a fixed background.  The scalarization bands in
$\eta/M^2$ correspond to regions bounded by parabolas in the
$(\eta, M)$ plane (shadowed regions in the left panel of
Fig.~\ref{fig:bh}).  The scalar field profiles of these solutions have
$n=0,1,\dots$ nodes (top-right panel of Fig.~\ref{fig:bh}),
corresponding to the order number of the scalarization band. A similar
ladder of excited states was observed for scalarized NSs in
scalar-tensor theory~\cite{Lima:2010na,Pani:2010vc}.
The normalized scalar charge\footnote{In other theories with a
Gauss–Bonnet coupling the scalar charge and the asymptotic value of
the coupling are related by $Q/M=2f_{,\vp}(\vp_\infty)/M^2$, and
this can lead to a bound on the coupling constant
(e.g.~\cite{Mignemi:1992nt,Kanti:1995vq,Sotiriou:2013qea}).  It
should be noted that there is no such relation for qsGB because
$f_{,\vp}(\vp_\infty)=0$.} $Q/M$ of these solutions is shown in the
bottom-right panel of Fig.~\ref{fig:bh} as a function of
$\eta/M^2$. This plot shows the values of $\eta$ admitting a
scalarized solution for each value of the BH mass.

\textit{Spontaneous scalarization and neutron stars.}
\begin{figure}[t]
\includegraphics[width=0.5\textwidth]{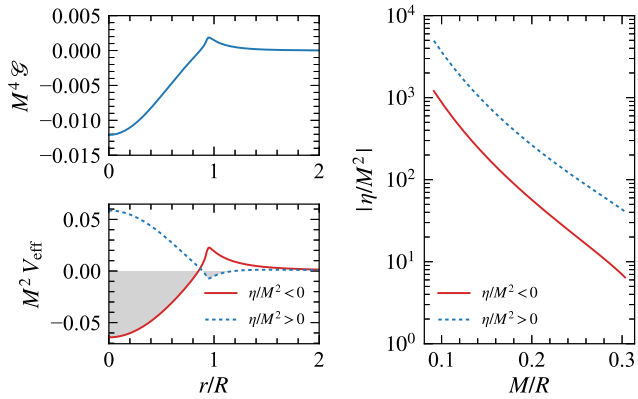}
\caption{{\it Tachyonic instability in a stellar background.} Left: profile
of the Gauss-Bonnet invariant (top) and of the effective potential (bottom),
inside a $M = 1.4$ M\msun~NS with the SLy4 EoS, assuming
$|\eta / M^2| = 100 / (1.4)^2 \sim 51$. The regions where the effective
potential becomes negative are shaded. Right: values of $\eta / M^2$ for
which the first bound state forms as a function of the compactness $M/R$.}
\label{fig:ns}
\end{figure}
Let us now consider NSs in qsGB gravity.  The Gauss-Bonnet invariant
for a static, spherically symmetric solution of the
Tolman-Oppenheimer-Volkoff (TOV) equations~\cite{Harrison:1965} is
\begin{equation}
\calg = \frac{48 m^2}{r^6} - \frac{128\pi(m + 2\pi r^3 p)\varepsilon}{r^3}\,,
\label{eq:GB_stars}
\end{equation}
where $m = r(1-1/b)/2$ is the mass function, and $p$ and $\varepsilon$
are the pressure and energy density inside the star, respectively. At
the surface $r=R$, $\varepsilon$ vanishes and~\eqref{eq:GB_stars}
matches smoothly the Schwarzschild value $\calg = 48 M^2 / r^6$, with
$M \equiv m(R)$ being the star's mass. We solve the TOV equations for
a ``canonical'' NS model with $M=1.4$ M$_{\odot}$, assuming the
SLy4~\cite{Douchin:2001sv} equation of state (EoS). The Gauss-Bonnet
invariant is mostly negative throughout the interior of the star (see
Fig.~\ref{fig:ns}, top-left panel); it is only positive near the
surface of the star, and in the exterior.  This suggests that if
$\eta < 0$, the scalar field can develop a tachyonic instability
inside the star, while if $\eta > 0$ the instability is triggered in
the outer region/exterior of the star.

In the bottom-left panel of Fig.~\ref{fig:ns} we show the effective
potential $V_{\rm eff}$ for the ``canonical'' NS model discussed
above, with $\eta = \pm 100$ M$^2_{\odot}$. As expected, there are
(shaded) regions where $V_{\rm eff}$ becomes negative. These regions
are inside the star when $\eta<0$, and outside the star when $\eta>0$.

Solving Eq.~\eqref{eq:waveeq_decoupl} in the NS background, we find
that scalarized solutions exist for both positive and negative values
of $\eta$. In the right panel of Fig.~\ref{fig:ns} we show the values
of $\eta/M^2$ corresponding to the lowest-lying scalarized solutions
with $\eta>0$ and $\eta<0$, as a function of the NS compactness.  Note
that scalarization occurs for lower values of $|\eta/M^2|$ when the
coupling constant is negative than when it is positive.

As in the BH case, we expect these results to translate into the
existence of scalarized NSs at the fully nonlinear
level~\cite{Pani:2011xm}, i.e. by integrating the modified TOV
equations obtained from
Eqs.~\eqref{eq:eom_scalar}-\eqref{eq:eom_metric} assuming a perfect
fluid for matter. Fully nonlinear stellar models will be explored in
forthcoming work.

\textit{Conclusions.}
We have identified and studied a subclass of scalar-tensor theories
with a coupling between the scalar and the Gauss--Bonnet invariant
that appears to exhibit spontaneous scalarization for both BHs and
NSs. Interestingly, BH scalarization does not have a single
threshold. Instead, for a given value of the coupling parameter $\eta$
hairy BHs exist when their mass lies in one of many narrow bands. Our
exploration for NSs strongly suggests that scalarization can take
place for both positive and negative values of $\eta$. However, the
effect appears to be stronger for negative values of $\eta$, for which
BH scalarization cannot occur. A full numerical study of NSs in these
theories is in progress and will be reported elsewhere.
It would be interesting to examine more closely the conditions under
which spontaneous scalarization can occur and its implications for the
structure of astrophysical BHs and compact stars, especially in binary
systems of interest for gravitational wave detectors.
A full study of the two-body problem in qsGB is beyond the
scope of this paper, but we anticipate interesting phenomenology
already at the post-Newtonian level~\cite{Yagi:2011xp}.
Binary systems containing scalarized BHs
and NSs (which have nonzero scalar charge $Q$) should emit dipolar
scalar radiation. However, in contrast with dilatonic and
shift-symmetric theories, where $Q \neq 0$ for {\it all} BHs, in our
case scalarization only happens -- and therefore dipolar radiation
would be emitted -- only in certain BH mass ranges (for a fixed
coupling $\eta$).
NSs in the shift-symmetric theory have
$Q = 0$~\cite{Barausse:2015wia,Yagi:2015oca}, thus evading the
stringent experimental constraints on dipolar radiation
emission from binary pulsars~\cite{Freire:2012mg}.
In qsGB gravity, if one of the NSs in the binary happens to be
scalarized, scalar radiation would be emitted, leaving a smoking gun
of the presence of the scalar field in the orbital dynamics.
It would also be interesting to investigate the strong field
dynamics of this theory. Apart from scalar-tensor
theories~\cite{Barausse:2012da,Berti:2013gfa,Shibata:2013pra,Palenzuela:2013hsa},
the application of numerical relativity simulations to other
theories of gravity is still in its
infancy~\cite{Benkel:2016kcq,Benkel:2016rlz,Okounkova:2017yby,Hirschmann:2017psw}.
To perform numerical simulations one must inevitably address the
issue of well-posedness~\cite{Papallo:2017qvl,Cayuso:2017iqc}, which
remains an open problem beyond the scope of our paper. By pointing
out the {\it existence} of potentially interesting phenomenology in
qsGB we hope to motivate further work in this direction.
Finally, it might also be worth extending our results to more
general couplings between the scalar field and the Gauss--Bonnet
invariant.

\textit{Note added.} Recently, a preprint studying a similar model
with BH spontaneous scalarization appeared in Ref.~\cite{Doneva:2017bvd}
and a study of evasions of no-hair theorems in sGB appeared in Ref.~\cite{Antoniou:2017acq}.

\textit{Acknowledgments.}~We thank Andrea~Maselli, Caio~F.~B.~Macedo,
Helvi~Witek, Paolo~Pani, Kent~Yagi and Nicol\'as~Yunes for numerous discussions.
This work was supported by the H2020-MSCA-RISE-2015 Grant No.
StronGrHEP-690904 and by the COST action  CA16104 ``GWverse''.
H.O.S was supported by NSF Grant No. PHY-1607130 and
NASA grants NNX16AB98G and 80NSSC17M0041.
J.S. was supported by funds provided to the Center for
Particle Cosmology by the University of Pennsylvania.
E.B. was supported by NSF Grants No. PHY-1607130
and AST-1716715.
T.P.S. received funding from the European Research Council
under the European Union's Seventh Framework Programme
(FP7/2007-2013) / ERC grant agreement n. 306425
``Challenging General Relativity''.
H.O.S also thanks the University of Nottingham for hospitality.

\bibliographystyle{apsrev4-1}
\bibliography{biblio}

\end{document}